\documentclass{ws-ijmpd}
\usepackage[super,compress]{cite}
\usepackage{hyperref}
\usepackage{mathtools}
\DeclareMathOperator{\sech}{sech}
\begin{document}


%
\catchline{}{}{}{}{}
%

\title{Phantom fluid supporting traversable wormholes in alternative gravity with extra material terms}

\author{P.K. Sahoo}

\address{$^{\dagger}$Department of Mathematics, Birla Institute of Technology and Science-Pilani, \\ Hyderabad Campus, Hyderabad-500078, India\\
pksahoo@hyderabad.bits-pilani.ac.in}

\author{P.H.R.S. Moraes}

\address{ITA - Instituto Tecnol\'ogico de Aeron\'autica - Departamento de F\'isica, 12228-900, S\~ao Jos\'e dos Campos, S\~ao Paulo, Brasil\\
Email: moraes.phrs@gmail.com}

\author{Parbati Sahoo}

\address{Department of Mathematics, Birla Institute of Technology and Science-Pilani, \\ Hyderabad Campus, Hyderabad-500078, India\\
sahooparbati1990@gmail.com}

\author{G. Ribeiro}

\address{UNESP - Universidade Estadual Paulista ``J\'ulio de Mesquita Filho'' - Departamento de F\'isica e Qu\'imica, 12516-410, Guaratinguet\'a, S\~ao Paulo, Brasil\\
Email: ribeiro.gabriel.fis@hotmail.com}



\maketitle

\begin{history}
\end{history}

\begin{abstract}

Wormholes are tunnels connecting different regions in space-time. They were obtained originally as a solution for Einstein's General Relativity theory and according to this theory they need to be filled by an exotic kind of anisotropic matter. In the present sense, by ``exotic matter'' we mean matter that does not satisfy the energy conditions. In this article we propose the modelling of wormholes within an alternative gravity theory that proposes an extra material (rather than geometrical) term in its gravitational action. Our solutions are obtained from well-known particular cases of the wormhole metric potentials, named redshift and shape functions, and yield the wormholes to be filled by a phantom fluid, that is, a fluid with equation of state parameter $\omega<-1$. In possession of the solutions for the wormhole material content, we also apply the energy conditions to them. The features of those are carefully discussed.

\end{abstract}



\tableofcontents

\section{Introduction}
 
In Einstein's General Relativity (GR), there is a kind of solution that describes geometrical bridges which can connect two far-off regions in the universe or even two separated universes. Those bridges are known as wormholes (WHs). The WH concept was firstly proposed by Hermann Weyl (1921) in Ref.\cite{Weyl/1921} in connection with mass analysis of the Einstein's field equations (although he did not use the term ``wormhole'' but ``one-dimensional tubes'' instead).   

The first WH solution discovered was the Schwarzchild WH (the simplest example of WH). Einstein and Rosen in 1935 transformed the radial coordinates of Schwarzchild metric to give this example of static WH which is also known as Einstein-Rosen bridge \cite{Einstein/1935}. The Schwarzchild WHs are considered as an intrinsic part of the maximally extended model of the Schwarzchild metric. Here, by ``maximally extended'' we mean the idea that space-time does not have any edges. 

A traversable WH can be visualized as a tunnel in space-time with two ends (or mouths), through which observers may safely traverse. The whole concept of traversable WHs is quite exciting because they admit the superluminal travel as a global effect of space-time topology. This was demonstrated by Morris and Thorne in a metric representing a static traversable WH \cite{morris/1988}.  

According to GR, traversable WHs are only possible if exotic matter exists at their throat, which involves an energy-momentum tensor violating the null energy condition (NEC) \cite{morris/1988}, which is in turn a part of the weak energy condition (WEC), whose physical meaning is that the energy density is non-negative in any reference frame.

On the other hand, modified gravity has been deeply analysed due to some GR apparent incompleteness in some regimes \cite{christodoulou/1981,alavirad/2013,capozziello/2008,geroch/1968,debono/2016}. The concept in these theories is basically to use an arbitrary but appropriate function in the gravitational action which generates extra terms in the field equations of GR. 

Once these alternative paths have shown good results (besides \cite{alavirad/2013,capozziello/2008}, check, for instance, \cite{Sepehri/2016,Yousaf/2017,Harko/2010,leon/2013,bamba/2013}), in this paper we will present the solution and behaviour of WHs under the $f(R,T)$ gravitational action \cite{harko/2011}, for which $R$ is the Ricci scalar and $T$ is the trace of the energy-momentum tensor. Using the appropriate metric for the static WH, the importance of the redshift and shape functions will be observed on different cases, analysing the possibility of generating a traversable WH.

It is important to quote that $f(R,T)$ gravity has been applied to different areas of astrophysics and cosmology, yielding interesting and observationally testable results, as it can be checked, for instance, in \cite{mcl/2017,amam/2016,clmaomm/2017,ms/2017,shabani/2014,alvarenga/2013}. 

Some more recent $f(R,T)$ gravity applications are presented in the following. Some analysis about compact stellar structures in $f(R,T)$ gravity were made \cite{sharif/2018,yousaf/2018}. Particularly, the hydrostatic equilibrium configurations of strange stars were obtained \cite{deb/2018,deb/2018b}, with the latter reference regarding an anisotropic distribution of matter inside such stars. Anisotropic stellar filaments evolving under expansion-free condition were analysed \cite{zubair/2018} and the dynamical stability of shearing viscous anisotropic fluid with cylindrical symmetry was investigated  \cite{azmat/2018}.

Perhaps the most intriguing of the observed features of the universe is its accelerated expansion \cite{Riess/1998,Perlmutter/1999}. While GR can explain it through the cosmological constant, which suffers from a serious fine-tuning problem \cite{weinberg/1989,hinshaw/2013}, alternative gravity can well address this particular issue \cite{akarsu/2014,alfaro/2013,akarsu/2015}. Also, a phantom fluid, that is, a fluid with equation of state (EoS) parameter $\omega<-1$, permeating the universe could well fit the observations indicating the cosmic acceleration \cite{gonzalez/2008,choudhury/2005,tonry/2003}.  

From some particular well addressed cases for the WHs redshift and shape functions, in this article we will obtain the material content solutions for WHs. In possession of those we will apply the energy conditions to them. Those disfavour a constant redshift function, which is broadly assumed in the literature. We will also derive the anisotropic dimensionless parameter (recall that WHs material content is described by an anisotropic energy-momentum tensor) and show that these WHs are filled and supported by a phantom fluid. The latter conclusion may yield some new thoughts on a cosmological scenario perspective and those are presented and discussed.

\section{An alternative gravity with extra material terms}

Nowadays the most popular alternative gravity theory is the $f(R)$ gravity \cite{sotiriou/2006,sotiriou/2010,de_felice/2010}, which takes general terms of $R$ in its gravitational action. Although being popular, it suffers from some persistent shortcomings \cite{amendola/2007,ganguly/2014,olmo/2007,chiba/2003}.

Instead of taking general terms of geometrical aspect in the action, one can also take material extra terms. That is the case, for instance, of the $f(R,\mathcal{L}_m)$ theory \cite{Harko/2010} and energy-momentum tensor squared gravity \cite{roshan/2016}.

Let us consider a gravity theory that takes terms proportional to the trace of the energy-momentum $T$ in its action. According to Ref.\cite{harko/2011}, such an action reads 

\begin{equation}\label{1}
S=\frac{1}{16\pi}\int d^{4}x\sqrt{-g}f(R,T)+\int d^{4}x\sqrt{-g}\mathcal{L}_m,
\end{equation}
where $f(R,T)$ is an arbitrary function of Ricci scalar, $R=R^i_j$ and $T=T^i_J$ is the trace of the stress-energy tensor of the matter with $g$ being the metric determinant. $\mathcal{L}_m$ is the matter Lagrangian density corresponding to the matter. Here, we have assumed the speed of light $c=1$ and gravitational constant $G=1$.

By varying the action $S$ with respect to the metric $g_{ij}$, the following $f(R,T)$ field equations are obtained \cite{harko/2011}
\begin{multline}\label{2}
R_{ij}f_R(R,T)-\frac{1}{2}f(R,T) g_{ij}+(g_{ij}\square-\nabla_i\nabla_j)f_R(R,T)=\\
8 \pi T_{ij}-f_T(R,T)\theta_{ij}-f_T(R,T)T_{ij}.
\end{multline}

Here, the notations are $f_R(R,T)=\partial f(R,T)/\partial R$, $f_T(R,T)=\partial f(R,T)/\partial T$ and the term $\theta_{ij}$ is defined as 
\begin{equation}\label{3}
\theta_{ij}=g^{\alpha \beta}\frac{\delta T_{\alpha \beta}}{\delta g^{ij}}=-2T_{ij}+g_{ij}\mathcal{L}_m-2g^{\alpha \beta}\frac{\partial^2 \mathcal{L}_m}{\partial g^{ij} \partial g^{\alpha \beta}}.
\end{equation}
Moreover, $R_{ij}$ and $g_{ij}$ are the Ricci and metric tensors.

We take, as it is is usually done in the literature, the stress energy momentum tensor of WHs as
  
\begin{equation}\label{4}
T_{ij}=(\rho+p_t)u_i u_j-p_t g_{ij}+(p_r-p_t)X_i X_j,
\end{equation}
where $\rho, p_r$ and $p_t$ are the energy density, radial pressure and tangential pressure, respectively. Moreover $u_i$ and $X_i$ are four velocity vector and radial unit four vector, satisfying the relations $u^{i} u_i=1$ and $X^i X_i=-1$.

It is important to highlight that $\mathcal{L}_m$ can be expressed with certain arbitrariness \cite{Harko/2010,Bertolami/2008}. We shall consider here that $\mathcal{L}_m=-\mathcal{P}$, with $\mathcal{P}=\frac{p_r+2p_t}{3}$ being the total pressure. In this way, Eq.(\ref{3}) can be rewritten as $\theta_{ij}=-2T_{ij}-\mathcal{P} g_{ij}$.

The $f(R,T)$ gravity field equations (\ref{2}) for $f(R,T)=R+2f(T)$ then takes the form
\begin{equation}\label{6}
R_{ij}-\frac{1}{2}Rg_{ij}=(8 \pi+2 \lambda) T_{ij}+\lambda g_{ij}(\rho-\mathcal{P}),
\end{equation}
where $f(T)=\lambda T$ and $\lambda$ is an arbitrary constant.

An interesting and intriguing property of the $f(R,T)$ theory of gravity, that can be extracted from Eq.(\ref{6}), is the non-conservation of the energy-momentum tensor. From \eqref{6},

\begin{equation}\label{6.1}
\nabla^{i}T_{ij}=\left(\frac{\lambda}{8\pi+\lambda}\right)\nabla^{i}g_{ij}(\mathcal{P}-\rho).
\end{equation} 

The particular consequences of the (non)-conservation of the energy-momentum tensor in $f(R,T)$ gravity have been explored. Inspired by \cite{josset/2017}, the cosmological consequences of the energy-momentum tensor non-conservation in $f(R,T)$ gravity were deeply investigated \cite{shabani/2017}. Also, the non-conservation of the energy-momentum tensor implies in non-geodesic motions for test particles in gravitational fields as it was deeply investigated \cite{baffou/2017}. In \cite{shabani/2018,mcr/2018} a different approach was considered. The authors have independently constructed a formalism in which an effective fluid is conserved in $f(R,T)$ gravity, rather than the usual energy-momentum tensor non-conservation.

\section{Wormhole field equations} 

The static spherically symmetric WH metric with Schwarzschild coordinates $(t,r,\theta, \phi)$ is \cite{morris/1988,visser/1995}

\begin{equation}\label{7}
ds^2=e^{a(r)}dt^2-\frac{dr^2}{1-\frac{b(r)}{r}}-r^2(d\theta^2+sin^2\theta d\phi^2),
\end{equation}
where $a(r)$ and $b(r)$ are the redshift function and shape function, respectively. Moreover, $r$ is the radial coordinate, which increases from a minimum radius value to $\infty$, i.e. $r_0\leq r <\infty$, where $r_0$ is known as the throat radius. A flaring out condition of the throat is considered as an important condition to have a typical WH solution, such that $\frac{b-b'r}{b^2}>0$  \cite{morris/1988} and at the throat, $r=r_0=b(r_0)$. Also, in order to have WH solutions, $b(r)$ must satisfy $b'(r_0)<1$. The absence of horizons and singularities is ensured when the redshift function $a(r)$ is finite and nonzero throughout the space time \cite{morris/1988}.

The main conditions for the shape function $b(r)$ are related to the shape of the WH, determined by the mathematics of embedding:
\begin{equation}\label{e1}
\frac{d^2r}{dz^2}=\frac{b(r)-b'(r)r}{2b^2(r)}>0
\end{equation}
at or near the throat. The function $z=z(r)$ determines the profile of the
embedding diagram of the WH,
\begin{equation}\label{e2}
z(r)=\pm \int_{r_0}^{r} \frac{dr}{\sqrt{\frac{r}{b(r)}-1}},
\end{equation}
which is obtained by rotating the graph of the function $z(r)$ around the vertical $z$-axis.

The general field equations (\ref{6}) for the metric (\ref{7}) are given as 
 \begin{equation}\label{8}
\frac{b(r)+a'r-a'r^2}{r^3}= -(8\pi + 2\lambda)p_r + \lambda\left(\rho - \frac{p_r+2p_t}{3}\right),
\end{equation}
\begin{multline}\label{9}
\frac{(2 r^2 a''+r^2 a'^2) (b-r)+r a' \left(r b'+b-2 r\right)+2 \left(r b'-b\right)}{4 r^3}\\=
-(8 \pi + 2\lambda)p_t + \lambda\left(\rho - \frac{p_r+2p_t}{3}\right),
\end{multline}
\begin{equation}\label{10}
\frac{b'(r)}{r^2}= (8\pi + 2\lambda)\rho + \lambda\left(\rho - \frac{p_r+2p_t}{3}\right).
\end{equation}

Also, by developing \eqref{6.1}, we obtain

\begin{equation}\label{10.1}
(8\pi+3\lambda)\rho'+\left(8\pi+\frac{5}{3}\lambda\right)(p_r'+2p_t')=0.
\end{equation}

From the above equations, the explicit form of the WH matter content, namely $\rho$, $p_r$ and $p_t$, are obtained as

 \begin{equation}\label{11}
 \rho=rF_1(r)\left(48\pi b'-\lambda\left(F_2(r)(b-r)+a'(F_3(r)+2)-16b'\right)\right),
 \end{equation}
 \begin{equation}\label{12}
 p_r=F_1(r) \left(48(r-1)ra'+\lambda r\left(-rF_2(r)+a'F_4(r)+8b'\right)+b\lambda(r(F_2(r)+a')-24)-48\pi b\right),
 \end{equation}
 \begin{equation}\label{13}
 p_t=\splitdfrac{F_1(r) (\lambda(2ra'+r(r(5F_2(r)+8a')-F_6(r))+b(12-5r(F_2(r)+a'))}{-12\pi\left(r(F_5(r)-F_2(r)-2a')+b(r(F_2(r)+a'')-2))\right)}.
 \end{equation}

In the above equations and henceforth a prime denotes derivative with respect to the radial coordinate $r$. Moreover, for mathematical simplifications, the functions $F_i(r)$, where $i$ runs from 1 to 6, were defined as
\begin{eqnarray}
F_1(r)\equiv\left(48(\lambda +2\pi)(\lambda +4\pi)r^3\right)^{-1}, \\
F_2(r)\equiv2ra''+ra'^2, \\
F_3(r)\equiv r(b'-4)+b, \\
F_4(r)\equiv r(b'+20)-22,\\
F_5(r)\equiv b(ra'+2), \\
F_6(r)\equiv b'(5ra'+4).
\end{eqnarray}

Furthermore, the dimensionless anisotropy parameter for anisotropic pressures, as the present case, is defined as \cite{Lobo/2013} 
\begin{equation}\label{an}
\Delta=\frac{p_t-p_r}{\rho}.
\end{equation}
Since $\rho >0$, the relation $\frac{\rho \Delta}{r}$ represents a force due to the anisotropic nature of the WH model. Geometry is attractive if $p_t<p_r$, i.e. $\Delta<0$, and repulsive if $p_t>p_r$, i.e. $\Delta>0$. The fluid is isotropic for $\Delta=0$, i.e. $p_r=p_t$.

\section{Wormhole Models with Hyperbolic Shape Function}\label{sec:whm}

\begin{figure}[h!]
  \includegraphics[width=75mm]{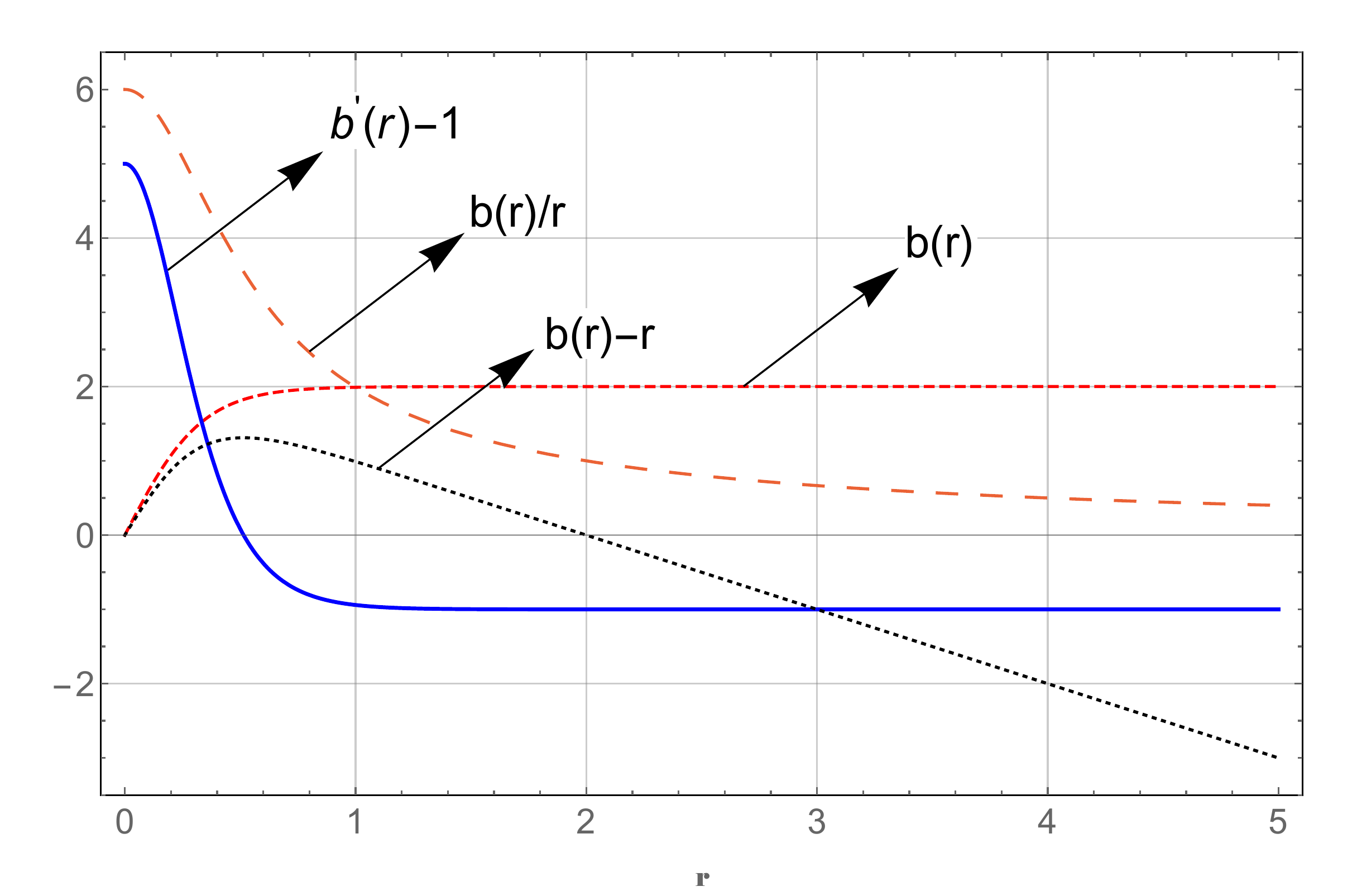}
  \caption{Variation of shape function with $m=2$ and $n=3$.}\label{fig1}
  \end{figure}

We will consider here the following specific form for the shape function \cite{Farook/2008}

 \begin{equation}\label{14}
 b(r)=m\tanh (nr),
 \end{equation}
where $m$ and $n>0$ are constants. As we quoted above, to admit the necessary metric conditions of WHs we have $b(r_0)=r_0$ and the flaring out condition $b'(r_0)-1<0$ as it is represented in the Fig.\ref{fig1}. 

From such a figure, we have $b(r)<r$ for $r>r_0$ and $\frac{b(r)}{r}\rightarrow 0$ as $r\rightarrow \infty$ showing an asymptotically flat behaviour. In the present article we have considered the WH throat where $b(r)-r$ cuts $r-$axis, i.e. $r_0=2$ as $b(2)\approx 1.99998$ with $m=2$ and $ n=3$.  
  
\subsection{Logarithmic redshift function}\label{ss:lrf}

In order to obtain traversable WHs, we consider a logarithmic form for the redshift function  as \cite{Pavlovic/2015}
 
\begin{equation}\label{15}
a(r)=\ln\left(\frac{r_0}{r}+1\right),
\end{equation} 

By substituting Equations (\ref{14}) and (\ref{15}) in (\ref{11}-\ref{13}), we obtain $\rho$, $p_r$, $p_t$ and the radial EoS parameter $\omega_r=\frac{p_r}{\rho}$ as

\begin{equation}\label{16}
\rho=\frac{F_1(r)}{(r+r_0)^2}\left[G_1(r) G_4(r)\sech^2(n r)-\lambda r_0(m \tanh(nr)(3r+2r_0)+G_7(r))\right],
\end{equation}
\begin{equation}\label{17}
p_r=\frac{F_1(r)}{(r+r_0)^2}\left[-m\tanh(nr)G_3(r)+\lambda G_1(r)(8r+7r_0)\sech^2(nr)+\lambda G_2(r)-48\pi r_0(r-1)(r+r_0)\right],
\end{equation}
\begin{equation}\label{18}
p_t=\frac{F_1(r)}{(r+r_0)^2}\left[m \tanh(nr)G_5(r)-G_1(r) \sech^2(nr)(\lambda(4r-r_0)+12\pi(2r+r_0))+G_6(r)\right],
\end{equation}
\begin{equation}
\omega_r(r)=\frac{-m \tanh(nr) G_3(r)+\lambda (G_1(r) (8r+7r_0)\sech^2(nr)+G_2(r))-48\pi r_0(r-1)(r+r_0)}{G_1(r) G_4(r)\sech^2(nr)-\lambda r_0(m \tanh(nr)(3r+2r_0)+G_7(r))},
\end{equation}

where $G_j(r)$, with $j$ running from 1 to 7, are expressed by the following forms

\begin{eqnarray}
G_1(r) & = & mnr (r+r_0),\\
G_2(r) & = & r_0(r(-24r-23r_0+22)+22r_0),\\
G_3(r) & = & \lambda(24r^2+4r r_0+22r_0^2)+48(r+r_0)^2,\\
G_4(r) & = & 48\pi (r+r_0)+\lambda(16r+17r_0),\\
G_5(r) & = & \lambda(12r^2+9r r_0+r_0^2)+12\pi r(2r+r_0),\\
G_6(r) & = & \lambda r_0(r(12r+7r_0-2)-2r_0)+12\pi r r_0(2r+r_0),\\
G_7(r) & = & r_0(r-2)-2r.
\end{eqnarray}

\subsubsection{Energy conditions}\label{sss:ec1}

The main energy conditions, such as the NEC, WEC, strong energy condition (SEC) and dominant energy condition (DEC) can be expressed directly in terms of $\rho, p_r, p_t$ as follows \cite{visser/1995,Hawking/1973,Wald/1984}:
 
 \begin{itemize}
\item WEC $\rightarrow$ $\rho \geq 0$,\,\,\ $\rho+p_i >0$,
\item NEC $\rightarrow$ $\rho+p_i \geq 0$,
\item SEC $\rightarrow$ $\rho+p_r+2p_t \geq 0$,
\item DEC $\rightarrow$ $\rho \geq \mid p_i \mid$.
 \end{itemize}
NEC represents the attractive nature of gravity. DEC states that the velocity of energy transfer cannot be higher than the speed of light. SEC stems from the attractive nature of gravity and its form is the direct result of considering a spherically symmetric metric in the GR framework. In a series of works, such as \cite{Harko/2013,Lobo/2007,Lobo/2008,Lobo/2009}, one can find that the energy conditions may be obtained to traversable WHs in modified gravity.

From the quantities above, we can plot the energy density as well as the energy conditions in Figs.\ref{fig3}-\ref{fig8} below. In all figures, we consider the free parameters $m=2$ and $n=3$.

\begin{figure}[h!] 
  \includegraphics[width=75mm]{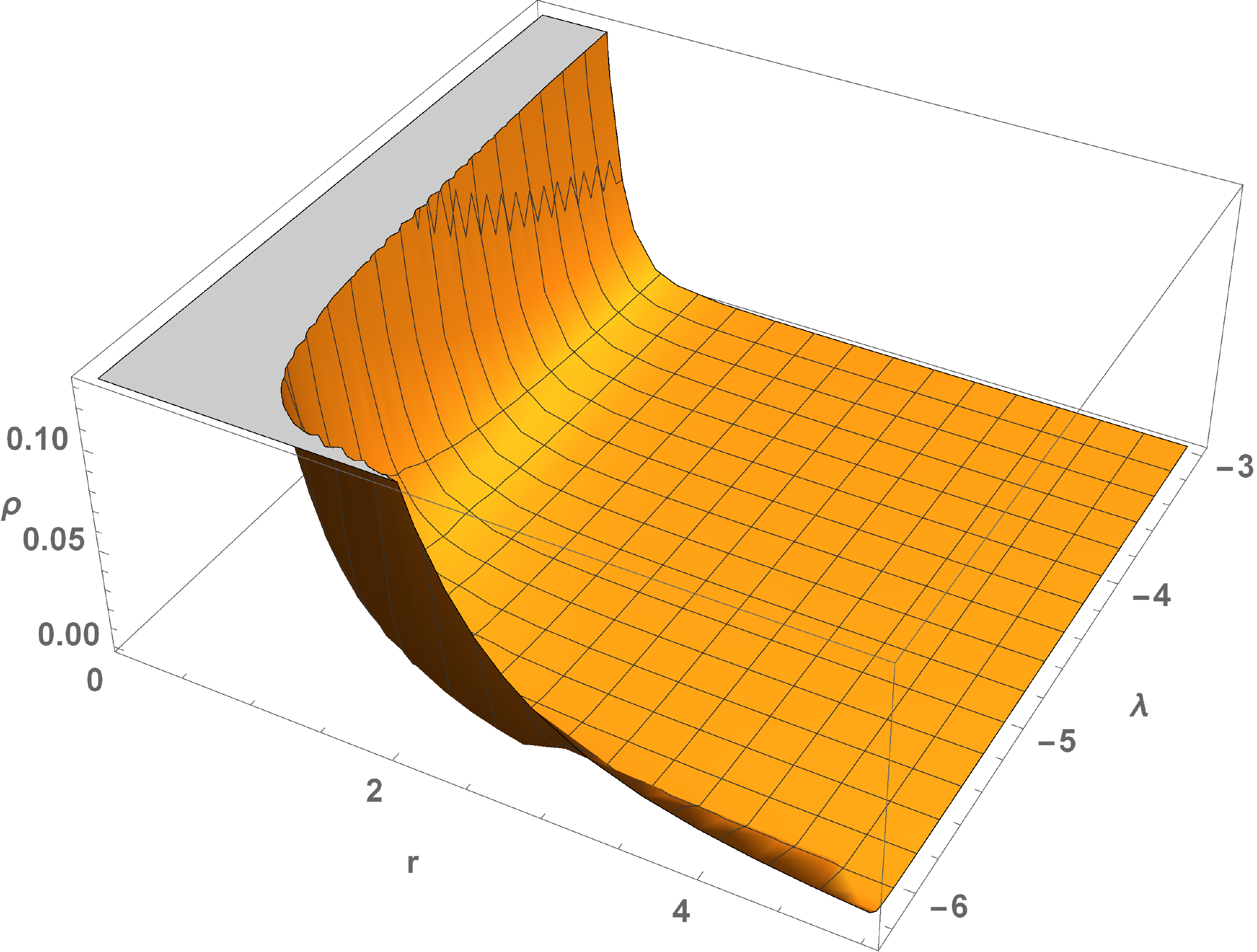}
  \caption{Energy density as a function of $r$ for different $\lambda$.}\label{fig3}
\end{figure}

\begin{figure}[h!]
  \includegraphics[width=75mm]{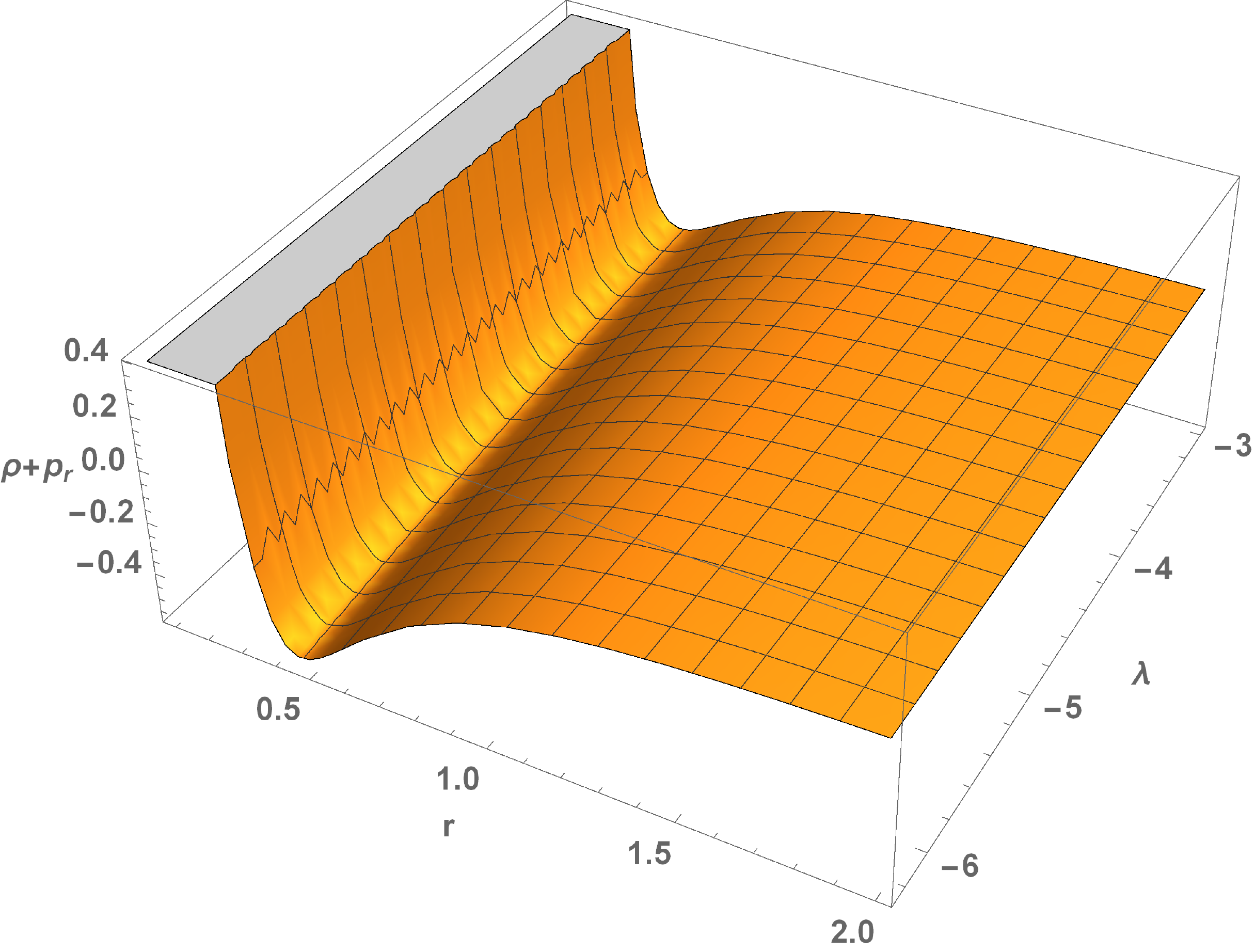}
  \caption{Null energy condition, $\rho+p_r\geq 0$.}\label{fig4}
 \end{figure}
 

From Fig.\ref{fig4} one can observe that NEC for $\rho+p_r$ validates for small $r$. Note that when NEC is violated, it implies that WEC will be also violated, while if WEC is valid, it does not imply that the NEC is satisfied. In the literature, these features have been discussed in different contexts including $f(R)$ gravity \cite{Lobo/2009,Bahamonde/2016}, $f(\mathcal{T})$ gravity \cite{Boehmer/2012,Jamil/2013}, with $\mathcal{T}$ being the torsion scalar, and curvature-matter coupling theories \cite{Garcia/2010}.

Fig.\ref{fig6} shows the validation of DEC for radial case. 

Moreover, in Figs.\ref{fig8}-\ref{fig9a} we plot the strong energy condition, radial EoS parameter and the dimensionless anisotropic parameter.

\begin{figure}[h!]
  \includegraphics[width=75mm]{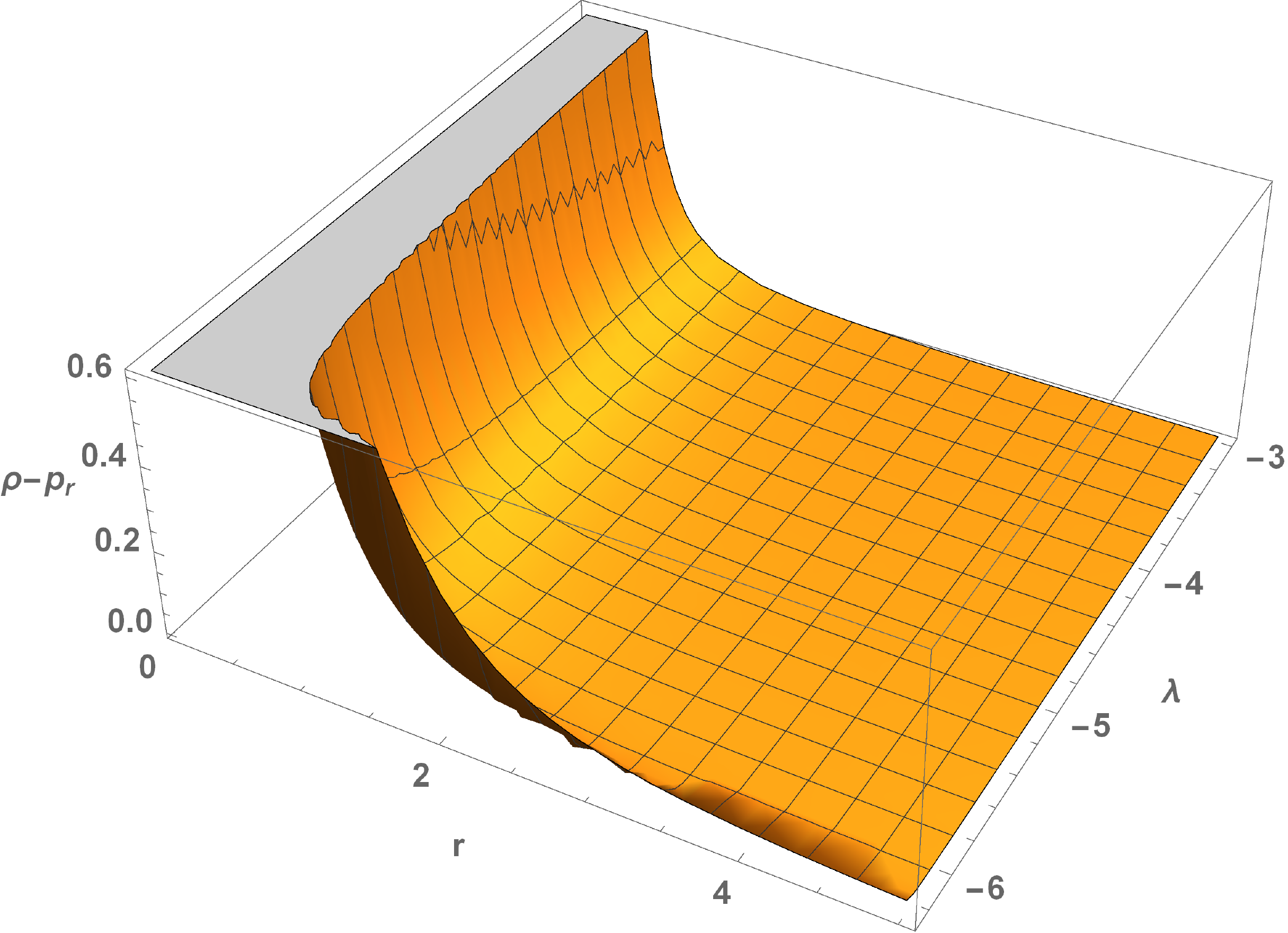}
  \caption{Dominant energy condition, $\rho\geq \vert p_r\vert$.}\label{fig6}
 \end{figure}

\begin{figure}[h!]
  \includegraphics[width=75mm]{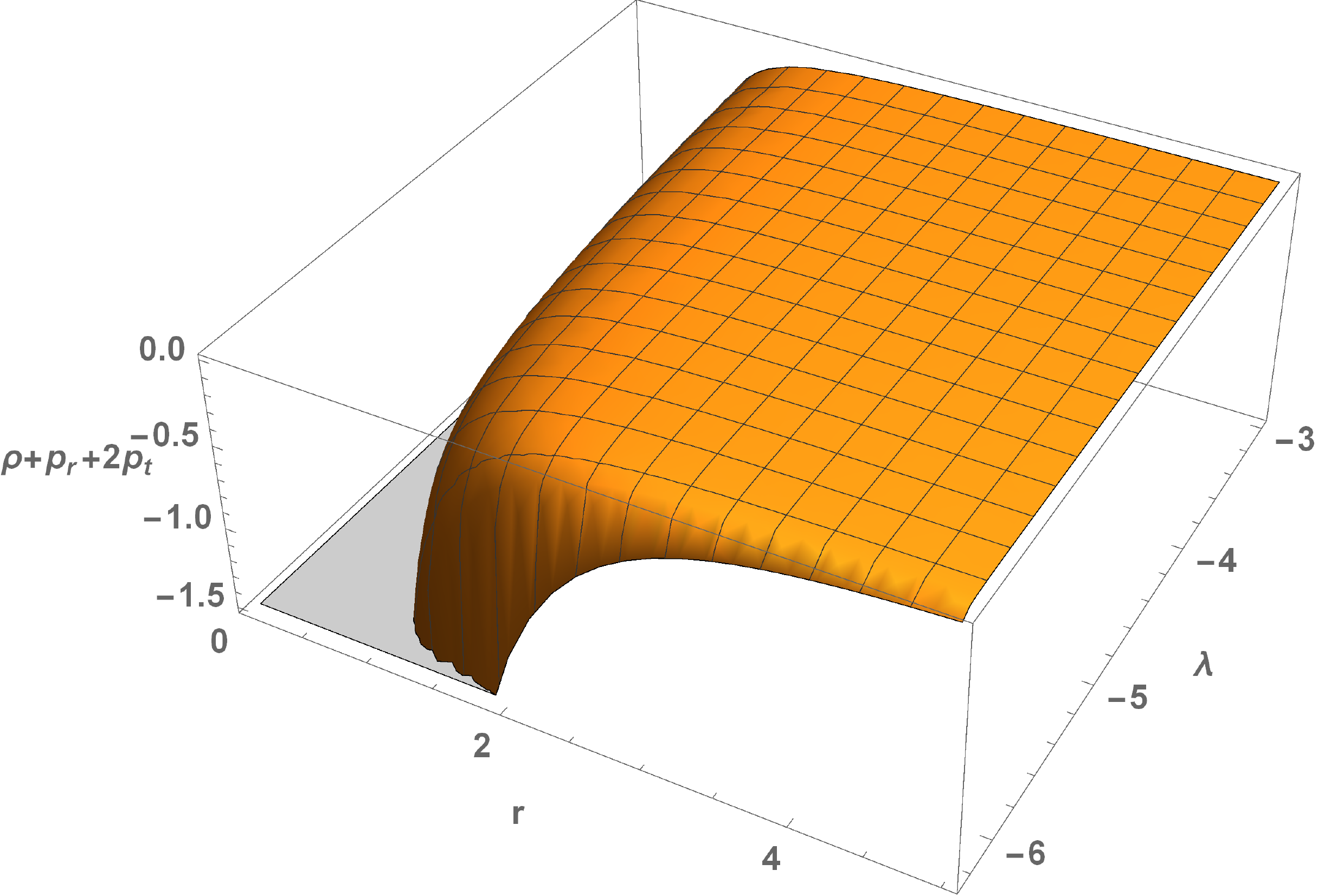}
  \caption{Strong energy condition, $\rho+p_r+2p_t\geq 0 $.}\label{fig8}
\end{figure}

\begin{figure}[h!] 
  \includegraphics[width=75mm]{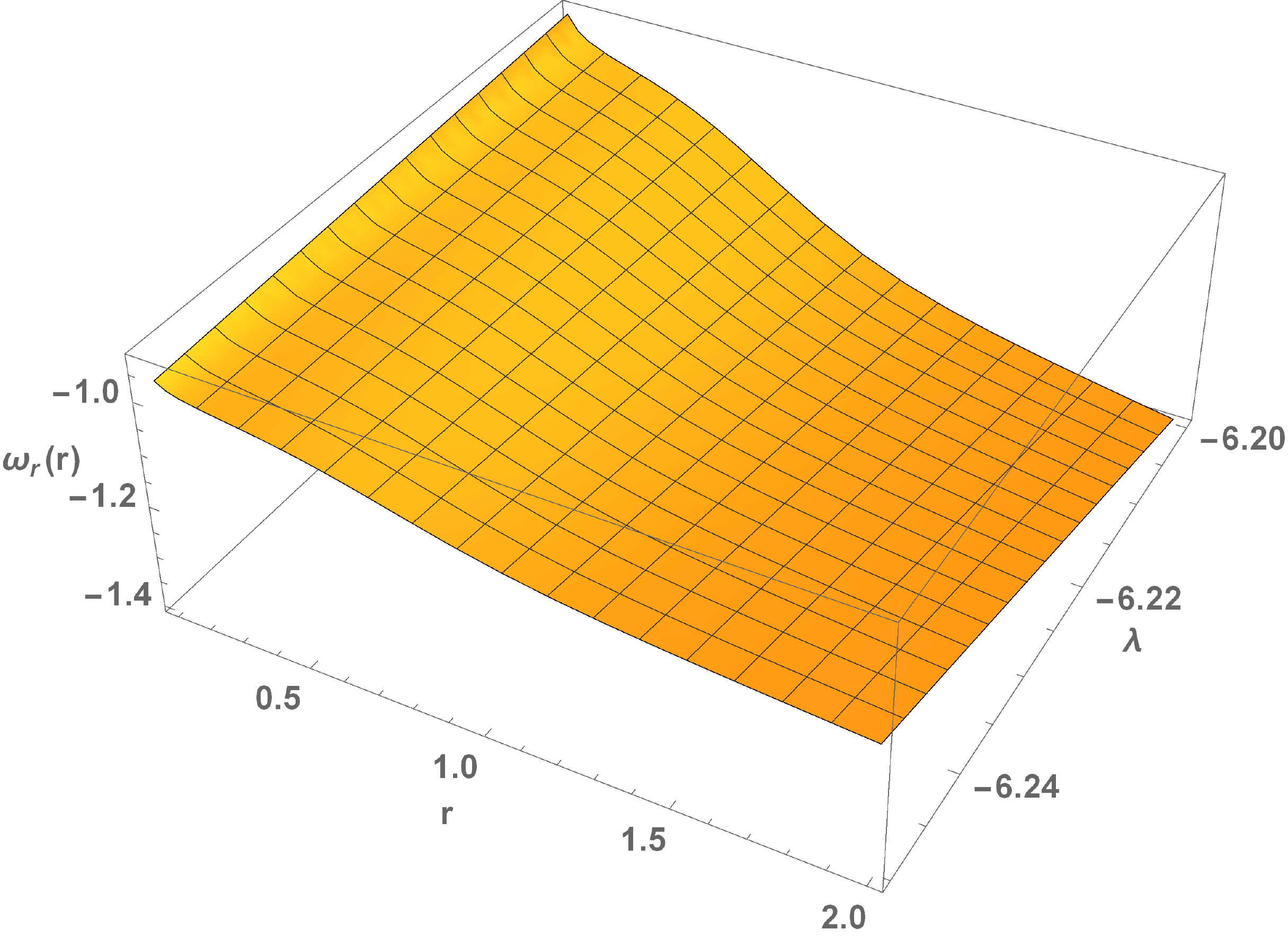}
  \caption{Radial equation of state parameter $\omega_r$ as a function of $r$ and $\lambda$.}\label{fig9}
\end{figure} 

\begin{figure}[h!] 
  \includegraphics[width=75mm]{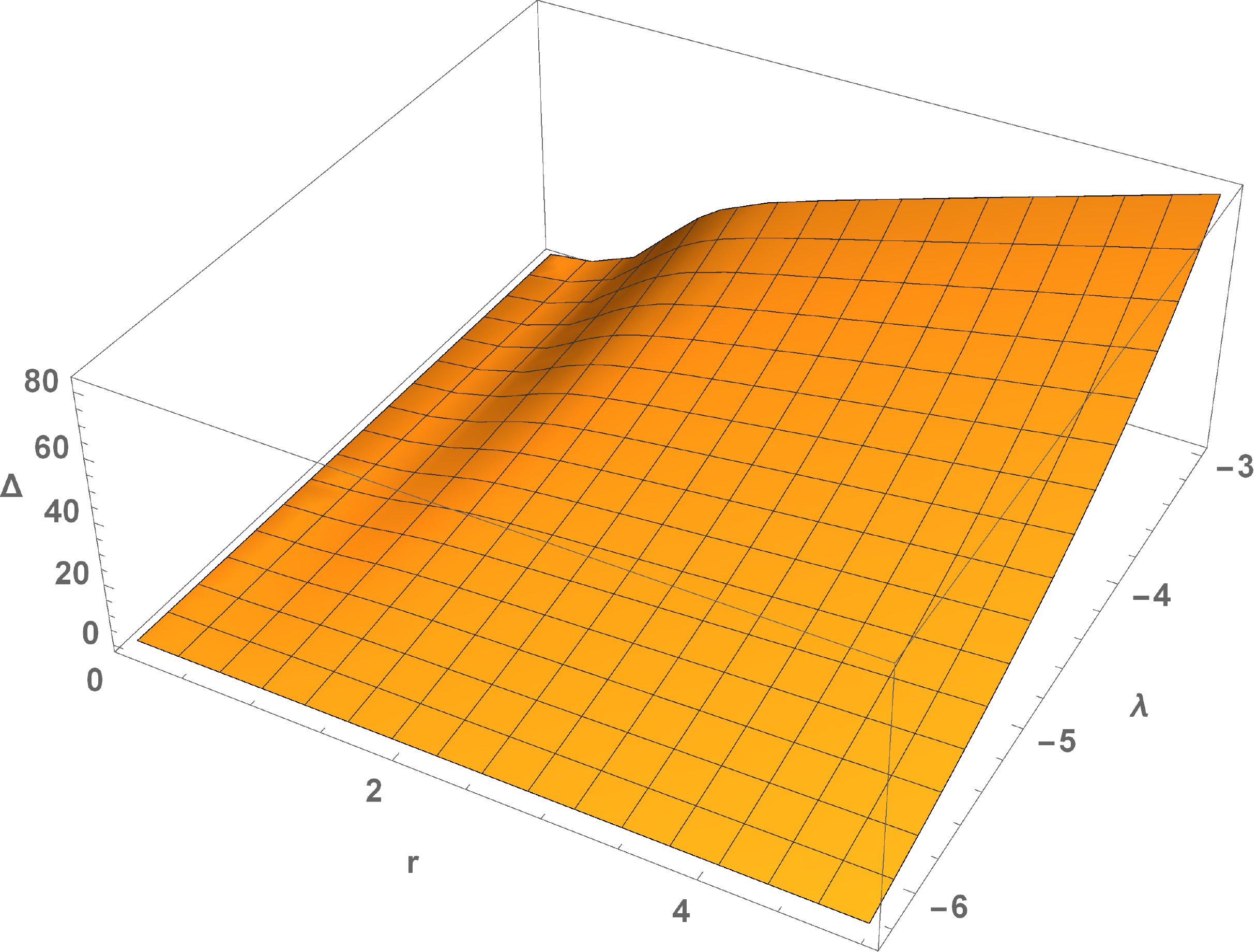}
  \caption{Dimensionless anisotropic parameter $\Delta$ as a function of $r$ and $\lambda$.}\label{fig9a}
\end{figure}

\subsection{Constant redshift function}\label{ss:crf}

In this case, we set the redshift function $a(r)=1$, such as in Refs.\cite{cataldo/2011,Farook/2007}, for example, as many others. Using this redshift function in Equations (\ref{11})-(\ref{13}), the explicit form of matter quantities reduces to

\begin{equation}\label{19}
\rho=\frac{rF_1(r)}{16} \left(3\pi+\lambda \right)b',
\end{equation}
\begin{equation}\label{20}
p_r=-\frac{F_1(r)}{8} \left[-\lambda r b'+(6\pi +3\lambda)b\right],
\end{equation}
\begin{equation}\label{21}
p_t=-\frac{F_1(r)}{4} \left[(6\pi +\lambda)rb'-(6\pi +3\lambda)b\right].
\end{equation}

The explicit expressions for the matter quantities and radial EoS parameter are obtained by substituting Eq.(\ref{14}) in the above equations, leading to

\begin{equation}\label{22}
\rho=\frac{rF_1(r)}{16}mn (\lambda +3 \pi ) \text{sech}^2(n r),
\end{equation}
\begin{equation}\label{23}
p_r=\frac{F_1(r)}{8} m \left[n \lambda  r \text{sech}^2(n r)-3 (\lambda +2 \pi ) \tanh (n r)\right],
\end{equation}
\begin{equation}\label{24}
p_t=\frac{F_1(r)}{2} m \text{sech}^2(n r) \left[3 (\lambda +2 \pi ) \sinh (2 n r)-2 n (\lambda +6 \pi ) r\right],
\end{equation}
\begin{equation}\label{25}
\omega_r(r)=\frac{2 \lambda  n r-3 (\lambda +2 \pi ) \sinh (2 n r)}{4 (\lambda +3 \pi ) n r}.
\end{equation}

\subsubsection{Energy conditions}\label{sss:ec2}

\begin{figure}[h!] 
  \includegraphics[width=75mm]{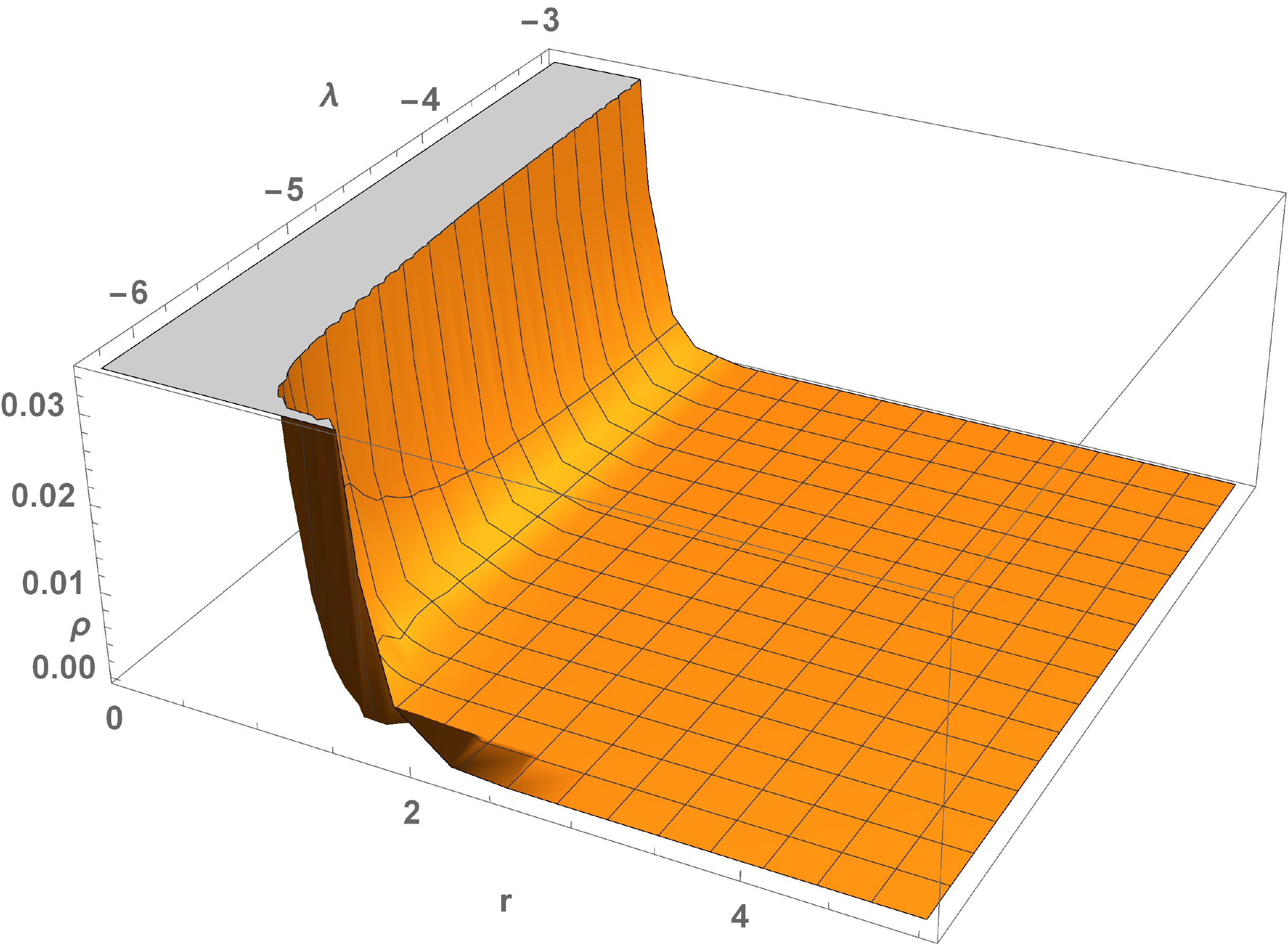}
  \caption{Variation of the energy density.}\label{fig10}
\end{figure}

\begin{figure}[h!]
  \includegraphics[width=75mm]{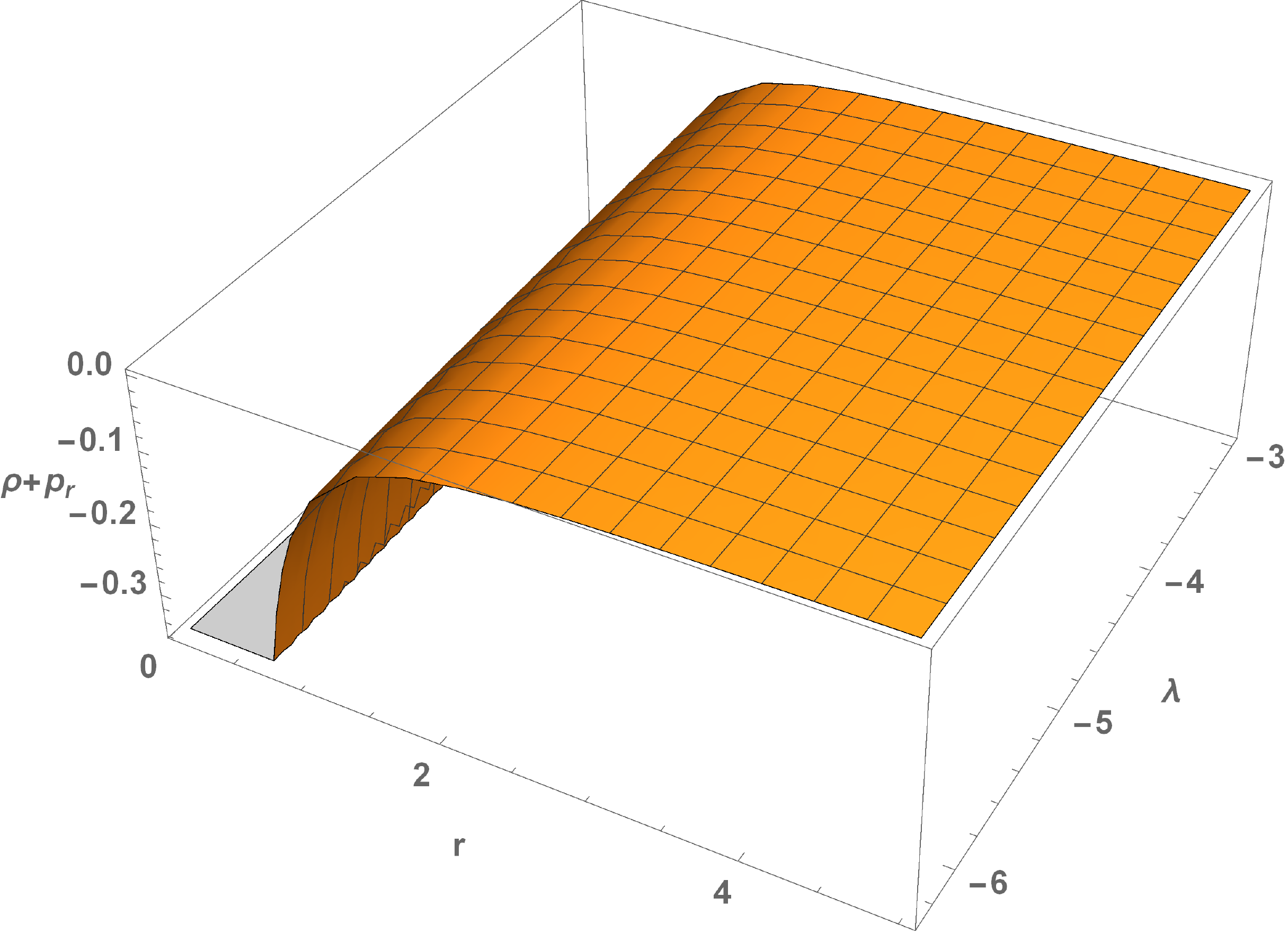}
  \caption{Null energy condition, $\rho+p_r\geq 0$.}\label{fig11}
 \end{figure}

\begin{figure}[h!]
  \includegraphics[width=75mm]{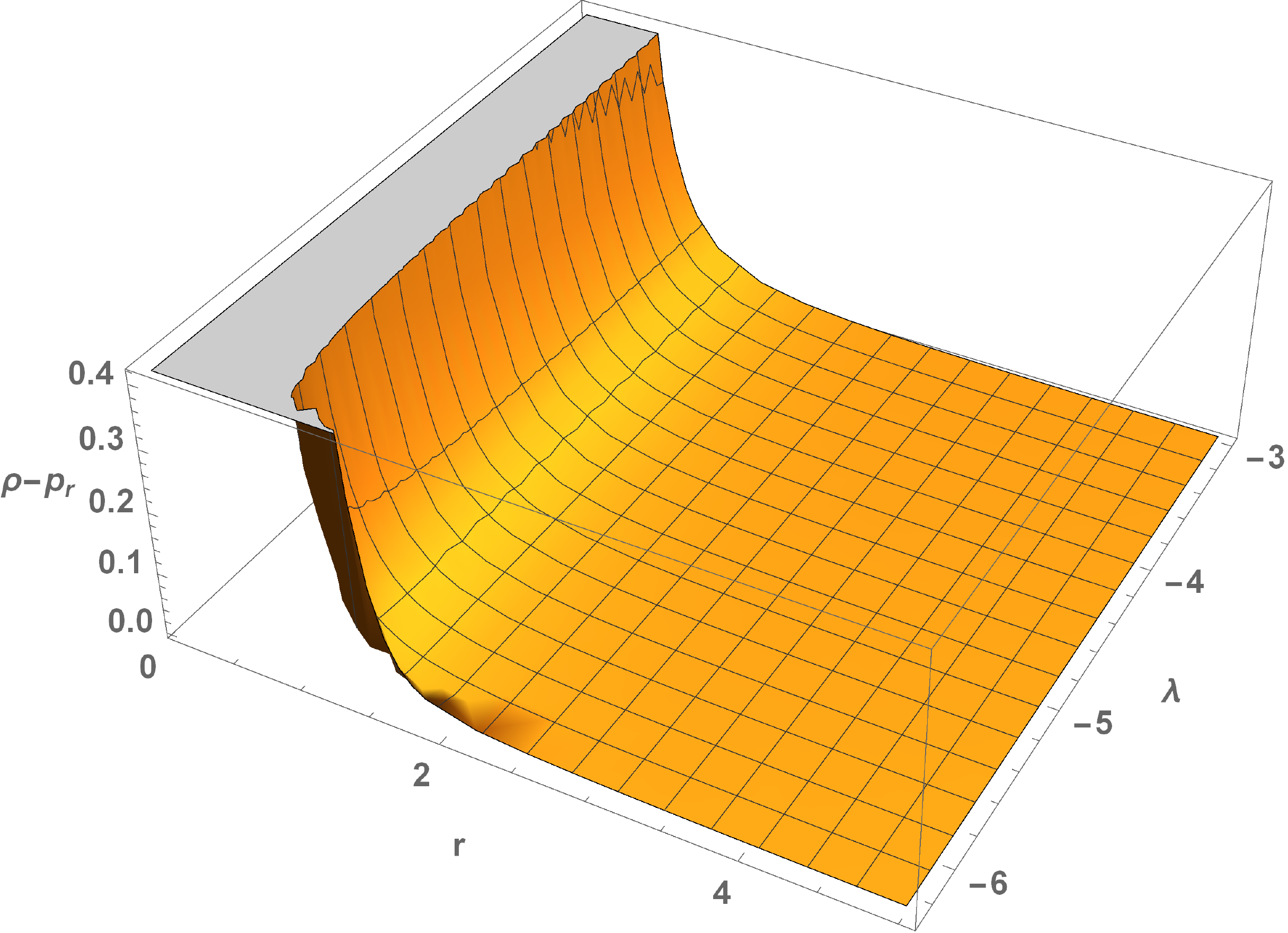}
  \caption{Dominant energy condition, $\rho\geq \vert p_r\vert$.}\label{fig13}
 \end{figure}

\begin{figure}[h!]
  \includegraphics[width=75mm]{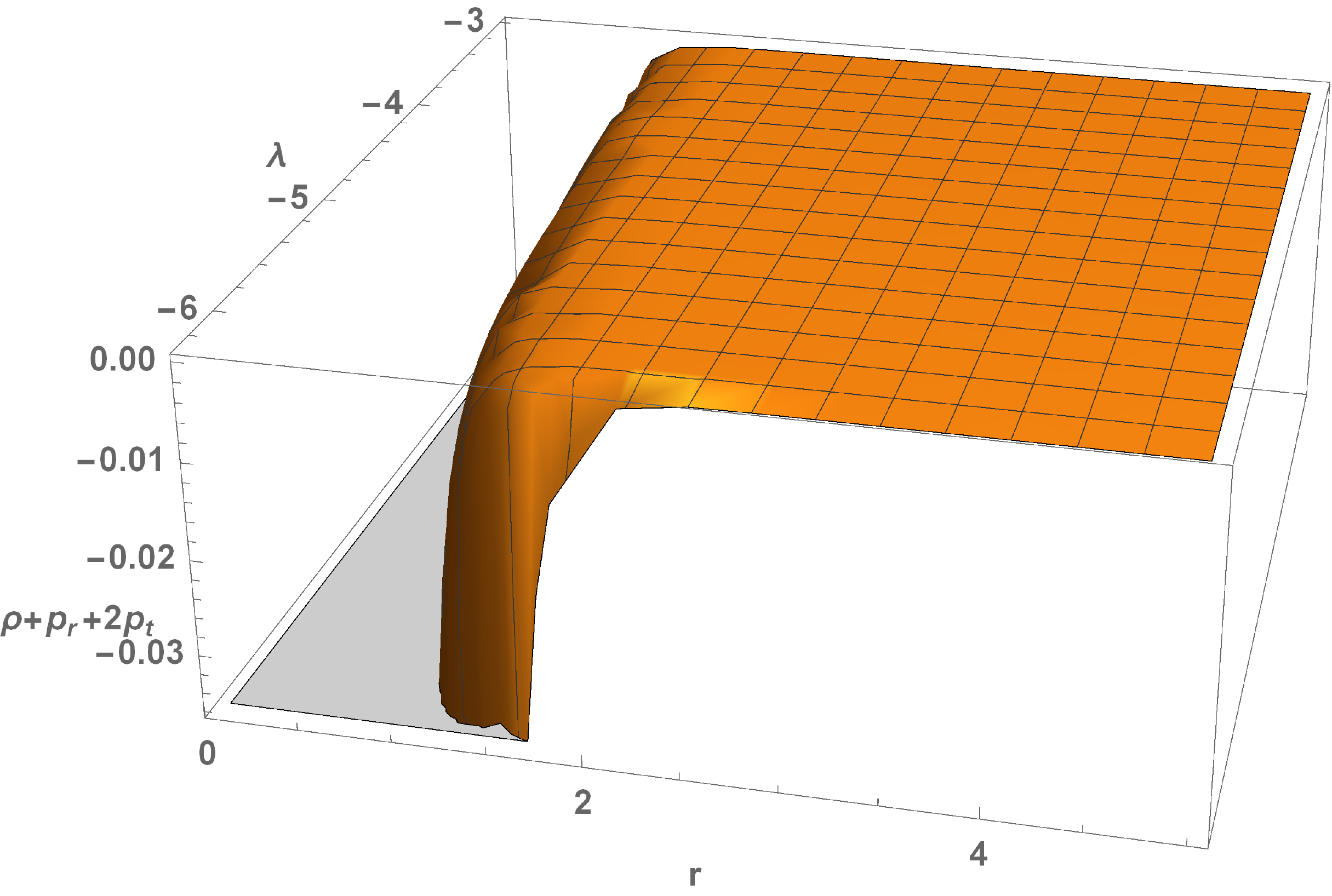}
  \caption{Strong energy condition, $\rho+p_r+2p_t\geq 0 $.}\label{fig15}
\end{figure}

From the quantities above, we plot the energy density as well as the energy conditions in Figs.\ref{fig10}-\ref{fig15}. In all figures we take $m=2$ and $n=3$.

The energy density is always positive as shown in Fig. \ref{fig10}.

From Fig.\ref{fig11} it is observed that NEC for radial pressure is violated.

DEC for radial pressure is valid for $r>0$ as plotted in Fig.\ref{fig13}. 

We can observe from Fig.\ref{fig15} that SEC violates everywhere and tends to zero for large values of $r$.

\begin{figure}[h!] 
  \includegraphics[width=75mm]{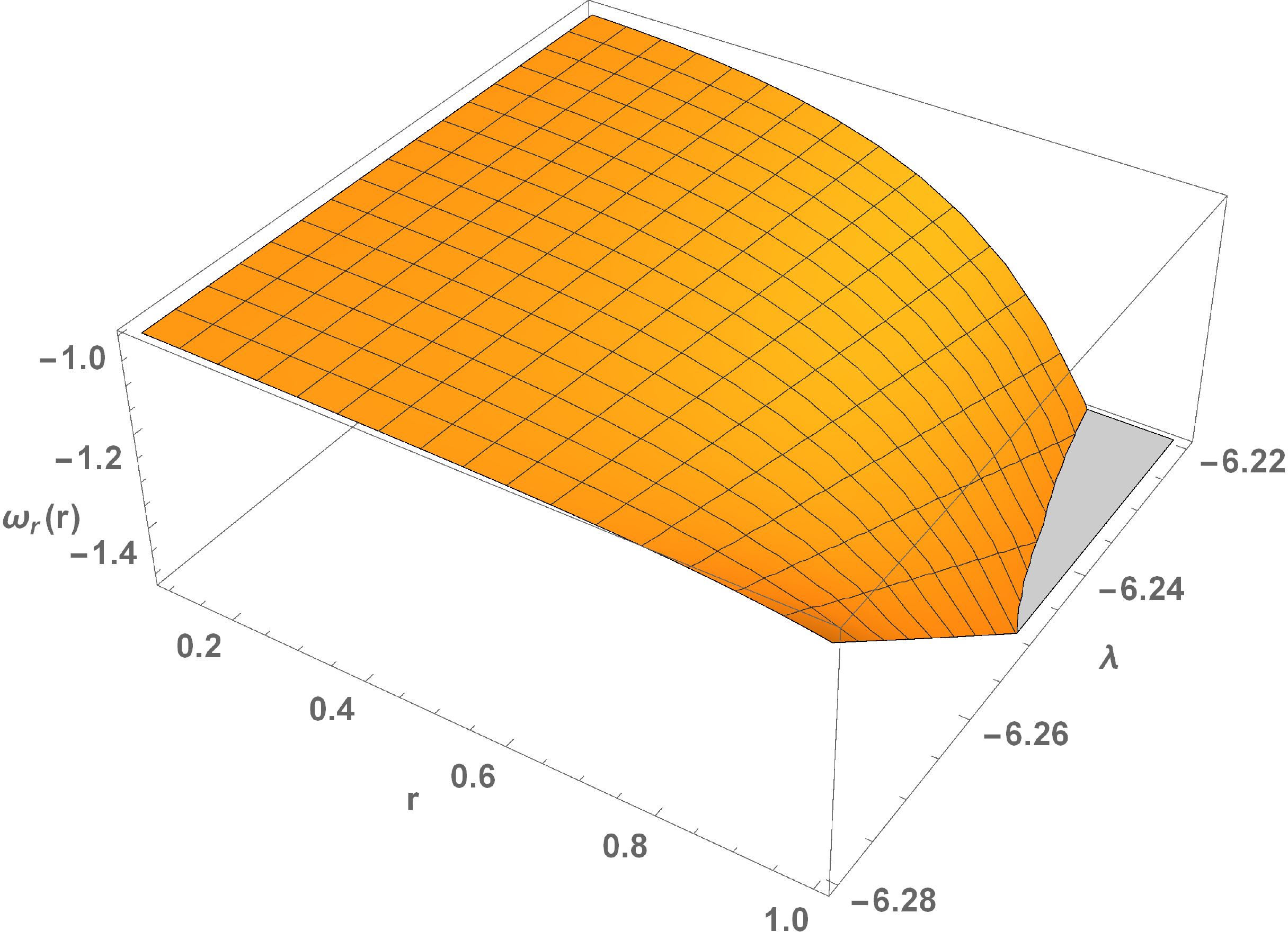}
  \caption{Radial equation of state parameter $\omega_r$ as a function of $r$ and $\lambda$.}\label{fig16}
\end{figure}

The radial state parameter $\omega_r(r)$ is always $<-1$ in support of the violation of NEC in radial pressure for modified gravity as plotted in Fig.\ref{fig16}.

\begin{figure}[h!] 
  \includegraphics[width=75mm]{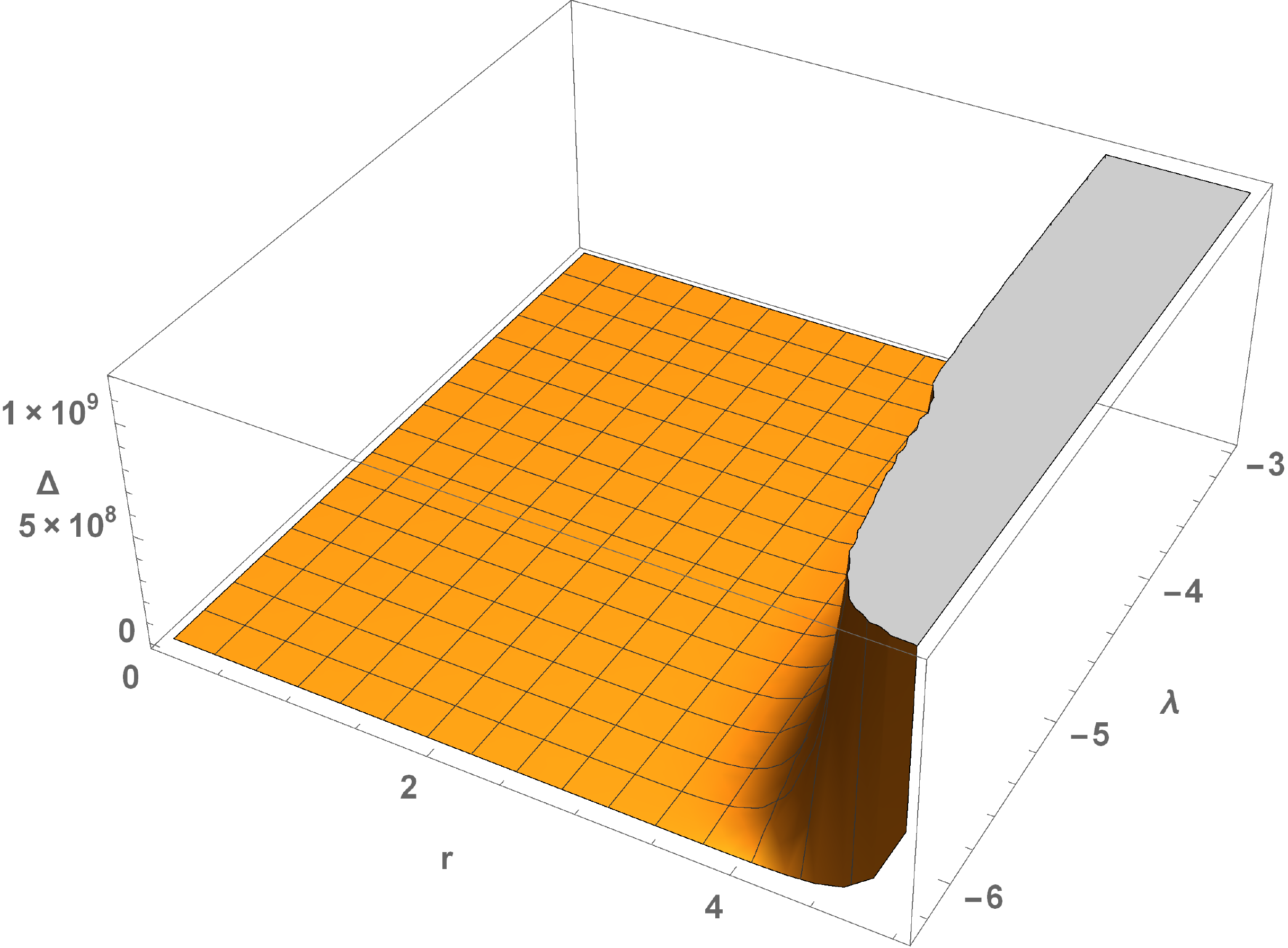}
  \caption{Dimensionless anisotropic parameter $\Delta$ as a function of $r$ and $\lambda$.}\label{fig17}
\end{figure}

We also plot the dimensionless anisotropic parameter in Fig.\ref{fig17}.

\section{Discussion}

We have proposed the $f(R,T)$ gravity as the background theory to deeply analyse WH solutions. The analysis of WHs in modified gravity theories is mainly occasioned by the possibility of obtaining material sector solutions that obey the energy conditions, so that they do not have to be referred to as ``exotic''. According to the $f(R,T)$ gravity authors, the $T$-dependence of the theory is motivated by the (possible) existence of imperfect fluids in the universe \cite{harko/2011}. In this way, the study of WHs, whose matter content is anisotropically distributed, in this theory is well motivated.

For the choice of the function $f(R,T)$, which {\it a priori} is arbitrary, we have taken $f(R,T)=R+2\lambda T$. The same choice has been made by many authors in many different applications of the theory, such as the recent references \cite{clmaomm/2017,sharif/2017,sharif/2017b,tiwari/2017,das/2017}.

We have then substituted the Morris-Thorne WH metric (\ref{7}) in the field equations (\ref{6}) of the model. The material solutions were obtained from a hyperbolic shape function, such as in Ref.\cite{Farook/2008}, for logarithmic \cite{Pavlovic/2015} and constant redshift functions \cite{cataldo/2011,Farook/2007}. 

Moreover, it is important to remark that departing from many references in the present literature, we have not assumed any EoS parameter. Rather, we have obtained it from the model. 

$\omega_r$ is plotted for the two cases in Figures \ref{fig9} and \ref{fig16}. There is a remarkable feature about $\omega_r$ that can be appreciated in those figures and will be discussed below.

In both figures, representing non-constant and constant redshift functions, the radial EoS parameter is $<-1$ for approximately the entire parameters space. This indicates that the concerned WHs are filled by a phantom fluid. Recall that a phantom fluid permeating the whole universe is an important alternative to explain the cosmic acceleration (besides the references in Introduction, check also \cite{srivastava/2005,elizalde/2009}).

Eq.\eqref{25} can be rewritten as a power series as follows:

\begin{equation}\label{fe}
\omega_r=-1-\frac{9 (\lambda +2 \pi ) r^2}{\lambda +3 \pi }-\frac{81 (\lambda +2 \pi ) r^4}{5 (\lambda +3 \pi )}+O\left(r^5\right),
\end{equation}
so that $\omega_r \rightarrow -1$ as $r\rightarrow 0$. The above equation explicitly shows the phantom aspect of the EoS obtained for the WH. As mentioned above, as $r\rightarrow0$, $\omega_r\rightarrow-1$. If that was the case for the whole WH, we would have a sort of ``dark energy wormhole''. However, as one gets away from $r=0$, the ``phantom contributions'' start to dominate and $\omega_r$ decreases its values, characterizing a phantom WH.

WHs filled by phantom fluids have been analysed in the literature as one can check Refs.\cite{Lobo/2013,lobo/2005,sushkov/2005,lobo/2005b,zaslavskii/2005,gonzalez/2009}. In the present article, a phantom fluid has shown to be responsible for supporting WHs with the geometrical features proposed in Section \ref{sec:whm} within $f(R,T)$ theory.

The results obtained from the energy conditions applications are quite interesting. Firstly, we have shown that it is indeed possible to respect the energy conditions in the present theory. Apart from SEC, all the energy conditions presented in Section \ref{sss:ec1}, from a non-constant redshift function, are respected, at least for a range of values of the radial coordinate.

On the other hand, the energy conditions shown in Section \ref{sss:ec2}, for a constant redshift function, have the NEC $\rho+p_r\geq0$ and SEC disobeyed. This may be an important clue that constant redshift functions yield unsatisfactory results regarding energy conditions applications, so that this particular case could be discarded from further WH modelling. 

The dimensionless anisotropic parameter given in Eq.(\ref{an}) is depicted in Figs.\ref{fig9a} and \ref{fig17}. This quantity was deeply approached in Ref.\cite{cattoen/2005}. In Ref.\cite{Lobo/2013}, an EoS was given in terms of $\Delta$. From Figs.\ref{fig9a} and \ref{fig17} it is clear that $\Delta>0$, which implies that the geometry is repulsive in both models due to the anisotropy of the system. One can conclude that, in principle, the repulsive character due to the anisotropy compensates the attractive nature of gravity for a range of the parameters of the WH models.

To finish, it is important to remark that in order to get the WH solutions we did not need to use Eq.(\ref{10.1}). The system of equations \eqref{8}-\eqref{10} has shown to be soluble from the assumptions in Section \ref{sec:whm}. In this way, in the present approach, the equation for the non-conservation of the energy-momentum tensor in $f(R,T)$ gravity merely puts a stricter bound in the values of the free parameter $\lambda$. In order to our solutions for $\rho$, $p_r$ and $p_t$ satisfy Eq.(\ref{10.1}), $-4\leq\lambda\leq-3$ in the logarithmic redshift function case (Section \ref{ss:lrf}) and $\lambda\sim-6.275$ in the constant redshift case (Section \ref{ss:crf}). Note that these values for $\lambda$ are in agreement with the energy conditions, as they should be.

{\bf Acknowledgements}\
PKS and PS acknowledges DST, New Delhi, India for providing facilities through DST-FIST lab, Department of Mathematics, where a part of this work was done. PHRSM would like to thank S\~ao Paulo Research Foundation (FAPESP), grant 2015/08476-0, for financial support.  The authors would also like to thank the anonymous referee for his/her important suggestions, that enriched the physical content of the paper.

\end{document}